\documentclass[iop]{emulateapj}%
\usepackage{amssymb}
\usepackage{epsfig}
\usepackage{apjfonts}
\usepackage{amsmath}
\usepackage{amsfonts}
\usepackage{graphicx}%
\providecommand{\U}[1]{\protect\rule{.1in}{.1in}}
\begin{document}

\title{ How far are the sources of IceCube neutrinos? Constraints from the diffuse
TeV gamma-ray background}
\author{Xiao-Chuan Chang \altaffilmark{1,2}, Ruo-Yu Liu\altaffilmark{1,3}, Xiang-Yu Wang\altaffilmark{1,2}}

\begin{abstract}
The nearly isotropic distribution of the TeV-PeV neutrinos
recently detected by IceCube suggests that they come from sources
at distance beyond our Galaxy, but how far they are is largely
unknown due to lack of any associations with known sources. In
this paper, we propose that the cumulative TeV gamma-ray emission
accompanying the production of neutrinos can be used to constrain
the distance of these neutrino sources, since the opacity of TeV
gamma rays due to absorption by the extragalactic background light
(EBL) depends on the distance that these TeV gamma rays have
travelled. As the diffuse  extragalactic TeV background measured
by \emph{Fermi} is much weaker than the expected cumulative flux
associated with IceCube neutrinos, the majority of IceCube
neutrinos, if their sources are transparent to TeV gamma rays,
must come from distances larger than the horizon of TeV gamma
rays. We find that above 80\% of the IceCube neutrinos should come
from sources at redshift $z>0.5$. Thus, the chance for finding
nearby sources correlated with IceCube neutrinos would be small.
We also find that, to explain the flux of neutrinos under the TeV
gamma-ray emission constraint, the redshift evolution of neutrino
source density must be at least as fast as the the cosmic
star-formation rate.

\end{abstract}

\affil{$^1$ School of Astronomy and Space Science, Nanjing University, Nanjing, 210093, China;  xywang@nju.edu.cn \\
$^2$ Key laboratory of Modern Astronomy and Astrophysics (Nanjing University), Ministry of Education, Nanjing 210093, China \\
$^3$ Max-Planck-Institut f\"ur Kernphysik, 69117 Heidelberg, Germany}


\section{Introduction}

The IceCube Collaboration recently announced the discovery of extraterrestrial
neutrinos \citep{Aartsen2013PhRvL,Aartsen2014PhRvL,Aartsen2015ApJ}. The sky
distribution of these events is consistent with isotropy
\citep{Aartsen2014PhRvL}. Such an isotropic distribution could be produced as
long as the distance of the source is significantly larger than the size of
the Galactic plane, thus extragalactic astrophysical objects are usually
proposed as their sources, but a Galactic halo origin is also possible
\citep{Taylor2014PhRvD}. The proposed extragalactic sources include galaxies
with intense star-formation
\citep{Loeb&Waxman2006JCAP,He2013,Murase2013PhRvD,Liu2014PhRvD,Chang&Wang2014ApJ,Tamborra2014JCAP,Chang2015ApJ,Chakraborty&Izaguirre2015PhLB,Senno2015ApJ,Bartos&Marka2015arXiv},
jets and/or cores of active galactic nuclei (AGNs)
\citep{Stecker1991PhRvL,Kalashev2014arXiv,Kimura2015ApJ,Padovani&Resconi2014MNRAS},
gamma-ray bursts
\citep{Waxman&Bahcall1997PhRvL,He2012,Liu&Wang2013Apj,Murase&Ioka2013PhRvL,Bustamante2015,Fraija2015arXiv}
and etc. The large uncertainty in our current knowledge about the distance of
neutrinos is due to that so far no associated astrophysical sources have been identified.

A common way to produce astrophysical high-energy neutrinos is the decay of
charged pions created in inelastic hadronuclear ($pp$) and/or photohadronic
($p\gamma$) processes of cosmic rays (CRs), in which high-energy gamma rays
will be also generated from the decay of synchronously created neutral pions.
The emissivity of gamma rays and neutrinos are related through $E_{\gamma
}Q_{\gamma}\left(  E_{\gamma}\right)  \approx\left(  2/3\right)  E_{\nu}%
Q_{\nu}\left(  E_{\nu}\right)  |_{E_{\nu}=E_{\gamma}/2}$ (for $pp$
process), where $Q$ represents the emission rate per source. VHE
gamma rays ($\gg 100\,$GeV), unlike neutrinos which propagate
through the universe almost freely, will be significantly absorbed
by the extragalactic background light (EBL) and cosmic microwave
background (CMB) during the propagation through intergalactic
space if the source distance is larger than the mean free path of
these gamma rays. For a TeV gamma-ray photon, the optical depth
would be larger than unity when the source locates at $z\ga0.1$.
Therefore, the cumulative flux of VHE gamma rays associated with
neutrinos carries the information about the distance of the
sources of these neutrinos.

Significant progresses have been made in our understanding of the
extragalactic gamma-ray background (EGB) in recent years. The
spectrum of the EGB has now been measured with the Fermi-LAT in
the energy range from 0.1 to 820 GeV \citep{Ackermann2015ApJ}. New
studies of the blazar source count distribution at gamma-ray
energies above 50 GeV place an upper limit on the residual
non-blazar component of the EGB \citep{Ackermann2015arXiv}. In
this paper, we use this upper limit at TeV energy to constrain the
distance and the evolution of the bulk population of neutrino
sources. The distance information has important implications for
the search of correlations between observed neutrino events and
nearby gamma-ray sources. If the inferred distances of the
majority of neutrino sources are large, {the search for nearby
correlated sources would be a challenge}. It may also explain the
negative result\footnote{See however, the result obtained by
\citet{Moharana&Razzaque2015}, who find that the arrival
directions of the cosmic neutrinos are correlated with $\ga$ 10
EeV UHECR arrival directions. This can be explained only if  the
sources are hidden  in TeV gamma-rays.} of search for the
correlation between neutrinos and ultra-high energy cosmic rays
obtained by \citep{Aartsen2015arXiv}, which originate within $\la$
100 Mpc.

In \S 2, we first present how we use the TeV emission to place constraints.
Then, in \S 3, we give the input conditions and assumptions. We give our
results in \S 4. Finally, we give the conclusions and discussions in \S 5.

\section{The method for constraints}

In the astrophysical origin scenarios, neutrinos (also gamma rays)
are produced in various discrete astrophysical objects, so the
total observed neutrino flux are the sum of contributions of each
individual source, rather than from truly diffuse emissions. We
assume that the neutrino sources are transparent to TeV gamma
rays, and only consider the attenuation in the intergalactic space
due to EBL and CMB absorption. {For simplicity, we assume here a
Poisson distribution for the closest sources\footnote{The
distribution of the closest galaxies  may be clumpy, which
suggests an overdensity of the nearby sources. Since the total
mass fraction in the nearby universe is small,  we find that this
overdensity hardly affects our results.}}, then the probability
density that the $n$th-closest source locates at a
comoving distance $r$ can be expressed as%
\begin{equation}
p\left(  n,r\right)  =\frac{4\pi N^{n-1}}{\left(  n-1\right)  !}e^{-N}%
r^{2}\rho_{0}%
\end{equation}
where $N$ is the expectation number of the sources within a
spherical comoving volume $V$ with radius $r$, i.e.
$N=\int\rho_{0}4\pi r^{2}dr$. So the expected comoving distance
where the $n$th-closest source locates is $\bar{r}\left( n\right)
=\int_{0}^{\infty}p\left(  n,r\right)  rdr$. {For distant sources,
the effect of fluctuation of distances are unimportant and the
distribution can be treated as a uniform distribution}, i.e.
$\rho=\rho\left( z\right)  $.

The influence of source number density on the gamma ray background
is complicated. If the spatial number density of the sources is
low, given a measured diffuse neutrino flux, the corresponding
pionic gamma ray luminosity of each source should be relatively
high, so this kind of sources are easier to be resolved by
instruments. By contrast, if the spatial number density of the
source is high, the luminosity of each source should be smaller.
As a result, these sources are more likely to be unresolved and
hence the emitted gamma rays contribute to the isotropic diffuse
gamma-ray background (IGRB). If more nearby sources are resolved
from background, the distant sources are allowed to be brighter
without violating the IGRB data. Such a requirement can be
expressed as,
\begin{equation}
\Phi_{\gamma,\text{un}}\left(  E_{\gamma}\right)  =\sum_{F_{n} <F_{\text{sens}%
}}^{n}F_{n}\left(  E_{\gamma}\right)  \leq\Phi_{\text{IGRB}
}\left( E_{\gamma}\right),
\end{equation}
where $\Phi_{\gamma,\text{un}}\left(  E_{\gamma}\right)  $ represents the
cumulative flux of unresolved sources, $F_{n}\left(  E_{\gamma}\right)  $
represents the flux of the $n$-th closest source and $F_{\text{sens}}$ is the
point source sensitivity of \textit{Fermi-}LAT.

In the same time, the resolved sources contribute to EGB together
with the unresolved ones, so they must satisfy
\begin{equation}
\Phi_{\gamma,\text{tot}}\left(  E_{\gamma}\right)
=\sum^{n}F_{n}\left( E_{\gamma}\right) \leq\Phi_{\text{EGB}}\left(
E_{\gamma}\right)                     ,
\end{equation}
{where $\Phi_{\gamma,\text{tot}}\left(  E_{\gamma}\right)  $ {
represents the cumulative flux of all sources, including both
resolved and unresolved ones. } In our calculation, we use the
broadband sensitivity provided by \textit{Fermi}-LAT performance
in Pass
8\footnote{http://www.slac.stanford.edu/exp/glast/groups/canda/\newline
lat\_Performance.htm}. Assuming the gamma-ray spectral index
$\gamma=2$, the sensitivity reaches a level of
$F_{\text{sens}}\sim2\times10^{-13}$ergs cm$^{-2}$s$^{-1}$.
According to this, we can determine whether a point source is
resolved or not. Based on the above two requirements (Eqs. 2 and
3), we will study how the maximum neutrino contribution from certain source is constrained by
the \textit{Fermi} data.}

For simplicity, we assume that all the sources have the same
intrinsic gamma-ray luminosity $L^{^{\prime}}$, which relates to
the gamma-ray spectral emissivity $Q_{\gamma}^{^{\prime}}\left(
E_{\gamma}^{^{\prime}}\right)  $ by
$L^{^{\prime}}=\int_{E^{\prime}_{\min}}^{E^{\prime}_{\max}}
Q_{\gamma }^{^{\prime}}\left(  E_{\gamma}^{^{\prime}}\right)
dE_{\gamma}^{\prime}$, where $E_{\mathrm{max}}^{\prime}$ and
$E_{\mathrm{min}}^{\prime}$ are the maximum and minimum energy of
emitted photons respectively. Here the prime denotes
quantities measured in the rest frame of the source (i.e., $E^{\prime}%
_{\gamma}=E_{\gamma}(1+z)$). While low-energy gamma rays can
propagate to us from the sources freely, VHE gamma rays may be
absorbed by the EBL and CMB photons in the intergalactic space.
The produced electron/positron pairs will also interact with EBL
and CMB photons and generate secondary gamma rays by
inverse-Compton scattering. Such a cycle is called cascade, and it
will continue until the newly generated photons are not energetic
enough to produce electron/positron pairs by interacting with the
background photons. As a result, the absorbed high-energy gamma
rays are reprocessed to a bunch of lower-energy ones. So the total
gamma ray flux after propagation consists of a primary component
which is the unabsorbed gamma rays, and a cascade component, i.e.,
\begin{equation}
F\left(  E_{\gamma}\right)  =\left\{  Q_{\gamma}^{^{\prime}}\left[  \left(
1+z\right)  E_{\gamma}\right]  e^{-\tau(E_{\gamma})}+Q_{\gamma,cas}\left(
E_{\gamma}\right)  \right\}  /4\pi\bar{r}^{2},
\end{equation}
where $\tau(E_{\gamma})$ is the optical depth for a photon of energy
$E_{\gamma}$. In this paper, we use the optical depth provided by
\citet{Finke2010ApJ}\footnote{http://www.phy.ohiou.edu/$\sim$%
finke/EBL/index.html} and discuss the effect of other EBL models
later.

Since the results depend on the density of the neutrino source, we
will consider three different cases: \newline1) High-density
source case, such as star-forming/starburst galaxies. The density
of starburst galaxies is about $4\times10^{-4}$Mpc$^{-3}$.
Starburst galaxies, due to their high star formation rates, and
hence large number of supernova or hypernova remnants therein, are
huge reservoirs of cosmic ray (CR) protons with energy up to EeV
\citep{Wang2007}. These CRs produce high-energy neutrinos by
colliding with gases in galaxies
\citep{Loeb&Waxman2006JCAP,Liu2014PhRvD}.
\newline2) Middle-density case, such as clusters of galaxies.
The density of clusters of galaxies is about
$\sim10^{-6}-10^{-5}$Mpc$^{-3}$ depending on the mass selection
\citep{Jenkins2001MNRAS}. Here we choose
$4\times10^{-6}$Mpc$^{-3}$ as a reference value. Galaxy clusters
have been argued to be able to accelerate CRs and considered as
possible sources for high-energy neutrinos \citep{Murase2008ApJ}.
\newline3) Low-density case, such as
blazars (e.g. BL Lacs and flat-spectrum radio quasars (FSRQs)).
Their density ranges  from $10^{-9}$Mpc$^{-3}$ to
$10^{-7}$Mpc$^{-3}$ \citep{Ajello2012ApJ,Ajello2014ApJ}. We choose
$4\times10^{-8}$Mpc$^{-3}$ as a reference value. As blazars are
powerful gamma-ray sources, there have been extensive discussions
about their possibility of being high-energy neutrino sources (see
\citet{Ahlers&Halzen2015RPPh} for a review).

{It should be noted that the three densities we chose are just
reference values to  study the effect of different source density.
Any specific sources should refer to the corresponding results
based on their spatial densities.}

\section{Assumption about the extragalactic neutrino flux and injection
spectra of gamma rays}

The latest combined maximum-likelihood analysis of IceCube neutrinos gives a
best-fit power law spectrum with a spectral index of $\gamma=2.50\pm0.09$ in
the energy range between 25 TeV and 2.8 PeV, and {an all-flavor flux} of
$\phi=\left(  6.7_{-1.2}^{+1.1}\right)  \cdot10^{-18}$GeV$^{-1}$s$^{-1}%
$sr$^{-1}$cm$^{-2}$ at 100 TeV \citep{Aartsen2015ApJ}. Interestingly, The
IceCube collaboration has tested the hypothesis of isotropy by analyzing data
in the northern and southern sky respectively. Compared to the all-sky result,
the spectrum of the events in the northern sky can be better fitted by a
harder power-law ($\gamma=2.0^{+0.3}_{-0.4}$), while the southern one favors a
slightly softer spectrum ($\gamma= 2.56 \pm0.12$). However, the result is not
conclusive, as the discrepancy could be simply caused by a statistical
fluctuation. Alternatively, it could be due to an additional component that is
present in only one of the hemispheres (either an unmodeled background
component, or e.g. a component from the inner Galaxy).

As indicated in some recent studies \citep{Ackermann2015arXiv},
the EGB above 50\thinspace GeV is
dominated by blazars at a level of $86_{-14}^{+16}%
\%$.  These are mostly low-luminosity hard-spectrum  BL Lacs. If
it is correct, this implies a strong suppression of contributions
from
other sources, which can contribute at most   a flux of $\la 2-3\times10^{-8}$GeV cm$^{-2}$s$^{-1}%
$sr$^{-1}$ at 50\thinspace GeV. If the neutrino sources are
transparent to gamma rays, the neutrino flux per-flavor is then
constrained to be at most $\sim10^{-8}$GeV
cm$^{-2}$s$^{-1}$sr$^{-1}$, which  can  explain the measured
neutrino flux at PeV energy, but is insufficient to explain the
flux at $\sim25$\thinspace TeV. To solve this tension, it has been
proposed that TeV neutrinos may come from some hidden sources,
i.e. they are not transparent to gamma rays
\citep{Murase2015arXiv,Bechtol2015arXiv}.

{Considering the above uncertainties, we divide our discussions
into two cases.  In the first case, we adopt the non-blazar EGB
obtained by \citet{Ackermann2015arXiv} to place constraints. We
assume that the gamma-ray background is only relevant to $\ga100$
TeV neutrinos whose sources are transparent to gamma-rays and
$\sim25$ TeV neutrinos may originate from some hidden sources or
from a component of the inner Galaxy. {Possible hidden sources
were suggested
\citep{Stecker1991PhRvL,Murase2015arXiv,Tamborra&Ando2015arXiv,Senno2015arXiv,Wang&Liu2015arXiv}
}}. The gamma-ray spectrum at the source should follow the
spectrum of neutrinos, which is assumed to be a flat spectrum with
index $\gamma=2.0$ below 1 PeV, as predicted by the Fermi
acceleration mechanism, and a steeper spectrum with $\gamma=2.5$
above 1 PeV \citep{Aartsen2014PhRvL}. Then the injection spectrum
of gamma rays can be expressed as
\begin{equation}
\label{spec_casea}\{%
\begin{array}
[c]{ll}%
Q_{\gamma}(E_{\gamma})\propto E_{\gamma}^{-2},E_{\gamma}<2(1+z)\mathrm{PeV} & \\
Q_{\gamma}(E_{\gamma})\propto
E_{\gamma}^{-2.5},E_{\gamma}\geq2(1+z)\mathrm{PeV}. &
\end{array}
\end{equation}
{In the second case, we relax this requirement by considering the
full EGB given in \citet{Ackermann2015ApJ}, allowing blazars to
contribute to IceCube neutrinos.  The neutrino spectrum at the
source is assumed to follow a broken power law with the spectral
index $\gamma=2$ at $E_{\nu}<25$ TeV and $\gamma=2.5$ at
$E_{\nu}>25$ TeV \citep{Aartsen2015ApJ}. Then the injection
spectrum of gamma rays can be expressed as }
\begin{equation}
\label{spec_caseb}\{%
\begin{array}
[c]{ll}%
Q_{\gamma}(E_{\gamma})\propto E_{\gamma}^{-2},E_{\gamma}<50(1+z)\mathrm{TeV} & \\
Q_{\gamma}(E_{\gamma})\propto E_{\gamma}^{-2.5},E_{\gamma}\geq50(1+z)\mathrm{TeV}. &
\end{array}
\end{equation}

\section{results}

When the cumulative TeV flux is fixed, as constrained by the IGRB
and EGB data, the total neutrino flux is affected by two factors.
First, as we already mentioned above, the spatial density of the
sources is an important factor. If the density is smaller, the
luminosity of individual source is larger and hence more nearby
sources will be resolved. Gamma rays from these resolved sources
will not be counted into IGRB while they still contribute to
diffuse neutrino flux. The second factor is the distance of the
source or the evolution of the source density with redshift. TeV
photons from more distant sources will be more likely to be
absorbed during propagation to the earth, while neutrinos will
not. So the neutrino flux will be higher if the fraction of
distant sources is higher (or the density evolution with redshift
is stronger). In our calculation, we adopt two forms of redshift
evolution for the purpose of illustration: one is the constant
density evolution, which means that the comoving density of the sources
does not change with redshift; the other one is the star formation rate (SFR)
evolution, for which the source density evolves as
$\varpropto(1+z)^{3.4}$ at $z<1$, $\varpropto(1+z)^{-0.3}$ at
$1<z<4$ and $\varpropto(1+z)^{-3.5}$ at $z>4$
\citep{Hopkins&Beacom2006ApJ,Yuksel2008ApJ}.

\subsection{Non-blazar EGB case}

In this case, we assume that EGB and IGRB are only relevant to
$\ga100$ TeV neutrinos and adopt the non-blazar EGB obtained by
\citet{Ackermann2015arXiv} as  an upper limit of the cumulative
gamma-ray flux from neutrino sources. First, we want to study what
fraction of the observed neutrino flux is contributed by nearby
sources. In Fig.~1, we fix the limit of gamma-ray flux to be
$2.5\times10^{-9}\mathrm{GeVcm^{-2}s^{-1}sr^{-1}}$ at
820\thinspace GeV, which corresponds to $14\%$ of the total EGB
(i.e., the non-blazar EGB\footnote{\citet{Ackermann2015arXiv} find
that blazars constitute about $86\%$ of the integrated photon flux
above 50\thinspace GeV. We here assume the spectrum of the summed
emission of all blazars are identical to that of the observed
EGB.}), and calculate the maximally allowed neutrino flux at PeV
as a function of the boundary distance $z_{\max}$. The boundary
distance $z_{\max}$ means that all the sources contributing to
IGRB/EGB locate within the redshift $z_{\max}$. The spectra of
gamma rays and neutrinos are assumed to follow
Eq.~(\ref{spec_casea}).

The left panel of Fig. 1 shows the maximally allowed neutrino flux
at 1\thinspace PeV without violating the non-blazar EGB. The
figure indicates that, in all three source density cases, only a
small fraction of neutrinos come from low redshift sources. The
sources below $z_{\max}=0.5$ can account for at most a fraction of
$20\%$ of the total neutrino flux. Thus, the majority of neutrinos
observed by IceCube should come from distance farther than
$z=0.5$. We can also see that, to account for the observed
neutrino flux, the redshift evolution of the sources should not be
slower than that of the cosmic star formation rate (SFR). The
constant evolution with respect to redshift can be ruled out. {The
above discussions only consider the constraints by the TeV
gamma-ray background. However, in the low source density case, the
nearest point sources could have been detected by IceCube, given a
sensitivity of $E^{2}dE/dN\sim10^{-12}$TeV s$^{-1}$cm$^{-2}$ for
IceCube \citep{IceCube2014b}. So we should also consider the
constraints by IceCube observation. This extra constraint suggests
that the source density should not be too low, unless part of
gamma ray emissions of the sources does not come from the same
hadronic process that produces neutrinos. Considering the latter
possibility for the low-density source case and the constraint by
IceCube non-detection of nearby sources, we re-calculate the
maximum neutrino flux, which is shown in the right panel of of
Fig.1. We find that the extra constraint affects the low density
case and the requirement of a fast evolution is strengthened.}

The requirement of a fast evolution of the source density with
redshift can also be seen by comparing the expected cumulative TeV
gamma-ray emission accompanying the production of neutrinos with
the observed gamma-ray background data. Fig. 2 shows the
cumulative gamma-ray emission for different redshift evolution
scenarios. Here we adopt the high source density case for
illustration, while the result holds for other source density
cases. We can see that a faster redshift evolution will lead to a
steeper gamma-ray spectrum. {Since \citet{Ackermann2015arXiv}
provide only the fraction of integral contribution to the total
EGB  above 50\,GeV by blazars, the exact fraction of non-blazar
EGB flux at 820 GeV is unclear. Thus, we draw a series of
horizontal lines in Fig.2 to show the different levels of the
non-blazar EGB fraction at 820 GeV.  If the non-blazar EGB flux is
lower than $\la 14\%$ of the total EGB flux at 820\,GeV, the
source evolution is required to be faster than that of SFR.} In
all three density evolution scenarios, however, the flux at 10-100
GeV are quite similar, which is naturally expected since 10-100
GeV gamma rays are almost as transparent as neutrinos. This
demonstrates the unique role of TeV gamma-ray flux in constraining
the evolution of neutrino source density.

\subsection{Full EGB case}

{In this case, we allow blazars to contribute to IceCube neutrinos
and use the full EGB as the upper limit \citep{Ackermann2015ApJ}.
The IGRB constraint becomes important for this case and we need to
consider this constraint as well. Thus, we fix the upper limits of
the cumulative gamma-ray flux at 820\thinspace GeV to be
$6\times10^{-9}\mathrm{GeVcm^{-2}s^{-1}sr^{-1}}$ for unresolved
sources (IGRB) and
$3\times10^{-8}\mathrm{GeVcm^{-2}s^{-1}sr^{-1}}$ for all the
sources (EGB) .}

{First, we assume that $10-100$ TeV neutrinos originate from
extragalactic sources that are transparent to TeV gamma rays.}
Fig. 3 shows the maximally allowed neutrino flux at {$\sim 25$}
TeV as a function of the boundary redshift $z_{\max}$, by assuming
a gamma-ray spectrum of Eq.~(\ref{spec_caseb}). The result
suggests a similar preference for high--redshift sources. In the
high and middle density cases, the source density evolution should
not be slower than the cosmic SFR evolution. {Since the luminosities
of these sources are relatively low, we find that IceCube
non-detection constraint  hardly affect the result. On the other
hand, for the low density case, the observed neutrino flux can be
achieved even for a constant source density evolution.  However,
when considering the neutrino non-detection constraint of nearby
point sources, only $\la 15\%$ of the EGB flux at 820 GeV can be
associated with neutrinos from the same hadronic process.
Including this extra constraint by IceCube non-detection, we find
that the source evolution should be faster than the SFR evolution for the
low-density case, as shown in the right panel of Fig. 3.} {Like in the
non-blazar case, we compare the cumulative gamma-ray emission of
unresolved sources with the IGRB data in Fig. 4 for the high
source density case. We find that, in order not to exceed the IGRB
at 820\,GeV, the source density must evolve faster than the SFR evolution}.

{However,  it has been suggested that the EGB above 50 GeV is
dominated by low-luminosity hard-spectrum BL Lacs
\citep{Ackermann2016ApJS}, which have a negative evolution with
redshift in source density \citep{Ajello2014ApJ}. If this is
correct, we can exclude the possibility that these BL Lacs are the
main sources of 10-100 TeV neutrinos observed by IceCube. This is
consistent with the conclusion in a previous study by summing up
neutrino flux from individual BL Lacs\citep{Padovani15}. {Thus,
the tension between the fact that the dominant sources producing
the gamma-ray background have negative evolution with redshift and
our conclusion that the sources of the high energy neutrinos must
have a fast, positive evolution with redshift argues for  hidden
sources for 10-100 TeV neutrinos.} Meanwhile, we should also note
that some types of blazars such as FSRQs, have a fast, positive
evolution with redshift and they only contribute a small part of
EGB/IGRB flux. These sources are still possible sources for
$\ga100$\,TeV IceCube neutrinos. This is consistent with the
recent discovery of a PeV neutrino which is in temporal and
positional coincidence with a high-fluence outburst from the FSRQ
PKS B1424-418 at redshift $z = 1.522$ \citep{Kadler2016}. For
these FSRQs, the situation is actually quite similar to the
low-density sources in the non-blazar case. }

\section{Discussions and conclusions}

In the above calculation, we used the EBL model given by
\citet{Finke2010ApJ}. Different EBL models might affect the
opacity of TeV gamma rays and we thus study this effect. We use
the upper and lower bounds on the opacity given by
\citet{Stecker2013arXiv} as the boundary of EBL uncertainties.
Fig. 5 shows the influence on cumulative gamma-ray emission when
varying the EBL opacity within this boundary. For illustration, we
adopt a local source density of $4\times10^{-4} \mathrm{Mpc^{-3}}$
with the SFR evolution. We find that, even for the strongest EBL
intensity model, the cumulative TeV flux decreases only slightly
(less than 20\%) compared to the case using the EBL model given by
\citet{Finke2010ApJ}. Thus, we conclude that the uncertainty from
EBL models does not change our conclusion that a fast source
density evolution is required.

In previous sections, we obtained the constraint on the source
density evolution assuming that the gamma-ray (or neutrino)
luminosity is the same for all sources. If the gamma-ray/neutrino
luminosity of the source also varies with redshift, it is then the
gamma-ray/neutrino emission rate density, rather than source
number density, that is constrained. The emission rate density can
be expressed as $g(z)\rho(z)$, where $\rho(z)$ represents the
source number density and $g(z)$ is the factor accounting for the
luminosity evolution of the source. As long as their product
$g(z)\rho(z)$, has a fast evolution with redshift, the requirement
is fulfilled. So the source number density may not need to evolve
that fast if the factor $g(z)$ evolves fast enough. The factor
$g(z)$ could originate from various physical causes. For example,
in the starburst galaxy scenario for IceCube neutrinos, cosmic
rays may have a higher pion production efficiency in
higher-redshift galaxies due to higher gas densities therein,
which leads to a larger $g(z)$ at higher redshifts
\citep{Chang2015ApJ}. {Besides, we note that, although
high-luminosity hard-spectrum BL Lacs  show only a mild evolution
in the redshift range of 1<z<2, they are severely deficient in
low-redshift (z <0.5) universe (Ajello et al. 2014).  These
sources can be essentially regarded as a fast evolution case, so
they may avoid the excess in the diffusive TeV gamma-ray
background when accounting for IceCube neutrinos.  }

To summarize, we find that extragalactic TeV gamma-ray background
is a useful tool to study the distance and density evolution of
neutrino sources. We have considered blazar and non-blazar source
models  and used different injection spectra correspondingly. In
both cases, we find that only a small fraction of neutrinos are
allowed to come from low-redshift sources in order not to exceed
the diffuse TeV gamma-ray background limit. To account for the
IceCube neutrino flux, the density of neutrino sources must have a
fast evolution with redshift. Interestingly, this is consistent
with the independent result obtained by the tomographic
constraints \citep{Ando2015PhRvL}.  {In addition, even for a fast
source density evolution, only $\ga100$ TeV neutrino flux could be
explained.} As our result shows that the IceCube neutrinos mainly
come from distant sources at high redshifts, any models arguing
for nearby sources may be ruled out as long as these sources are
transparent to TeV gamma rays. {Also, any search for nearby
sources correlated with IceCube neutrinos would face a challenge.}
Instead, we suggest to search for correlations with potential
cosmic ray accelerators at high redshifts.

\acknowledgments We thank Paolo Padovani for the discussions and
the anonymous referee for the invaluable report. This work is
supported by the 973 program under grant 2014CB845800, the NSFC
under grants 11273016, and the Excellent Youth Foundation of
Jiangsu Province (BK2012011).

\clearpage

\begin{center}

\begin{figure}[ptb]
\includegraphics[width=1.0\textwidth]{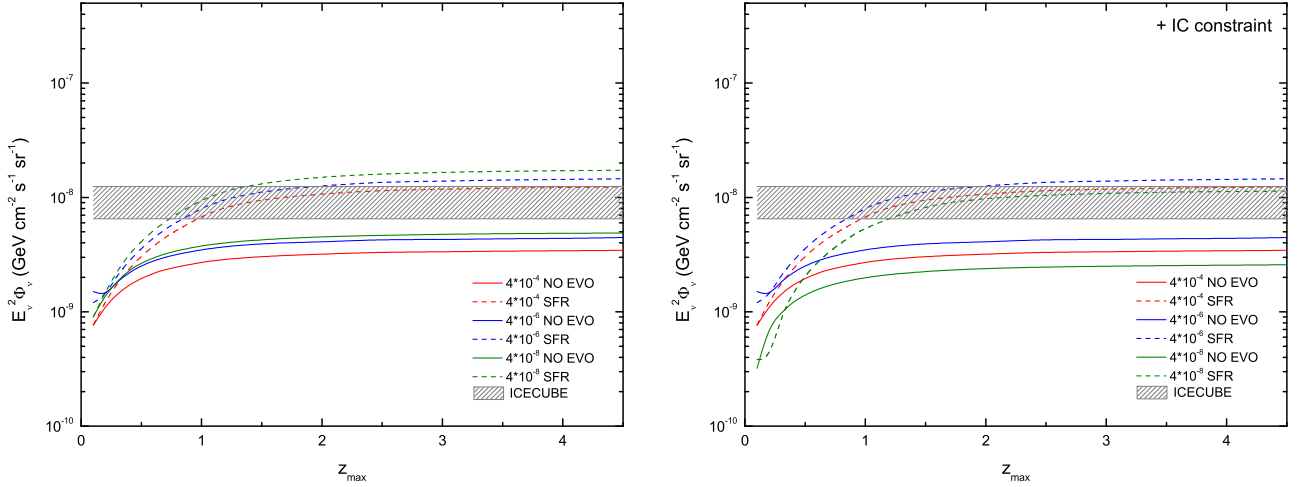}\caption{The maximally-allowed
neutrino flux at 1 PeV under the constraint of the non-blazar TeV
gamma-ray background as a function of the boundary redshift
$z_{\max}$ in the non-blazar source case. The red, blue and
green lines represent the cases with source density of $4\times10^{-4}%
$Mpc$^{-3}$, $4\times10^{-6}$Mpc$^{-3}$ and
$4\times10^{-8}$Mpc$^{-3}$, respectively. The solid and dashed
lines represent, respectively, no evolution and SFR evolution of
the source density with redshift. The grey shaded area represents
the observed neutrino flux (per flavor) by IceCube at 1 PeV with
errors \citep{Aartsen2014PhRvL}. In the right panel, an extra
constraint due
to the IceCube non-detection of nearest point sources is considered. }%
\end{figure}

\begin{figure}[ptb]
\includegraphics[width=0.9\textwidth]{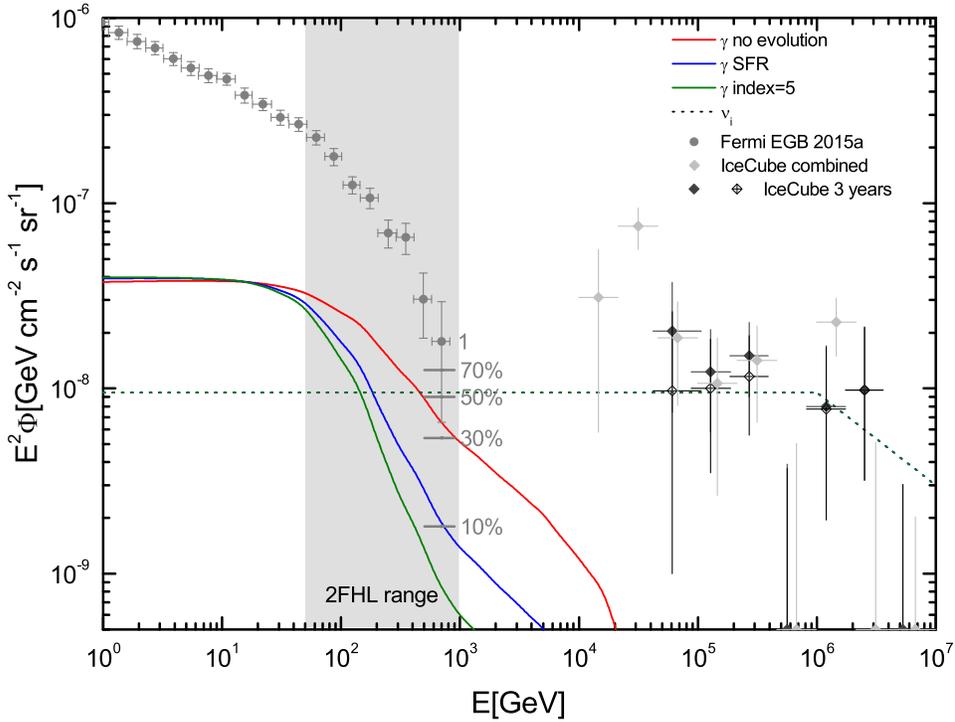}\caption{The cumulative gamma-ray
emission associated with production of IceCube neutrinos in
non-blazar case for different redshift evolution models. The
spectrum of the neutrinos used in the calculation is shown by the
short dashed line. The red, blue and green lines show,
respectively, the cases assuming no redshift evolution, SFR
evolution and an evolution given by $\varpropto(1+z)^{5}$ at $z<1\
$, $\varpropto
(1+z)^{\text{-0.3}}$ at $1<z<4$, and $\varpropto
(1+z)^{\text{-3.5}}$ at $z>4$. The IGRB data from \textit{Fermi}-LAT are
denoted by grey dots and the horizontal short grey lines shows 10\%, 30\%, 50\% and 70\% of EGB flux at 820 GeV, respectively
\citep{Ackermann2015ApJ}. The IceCube data are denoted by black and grey dots \citep{Aartsen2014PhRvL, Aartsen2015ApJ}.
In the calculation, we use the case of source density of $4\times10^{-4} \mathrm{Mpc^{-3}}$.}%
\end{figure}

\begin{figure}[ptb]
\includegraphics[width=1.0\textwidth]{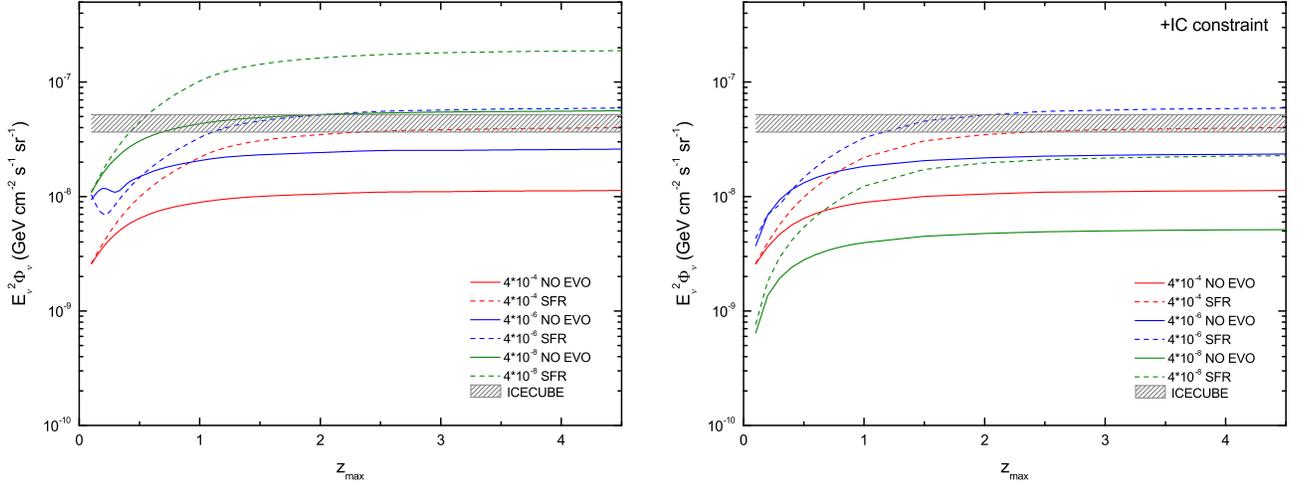}\caption{Same as Figure 1, but
for the full EGB case. Note that all the neutrino fluxes here
correspond to the values at the reference energy of 25 TeV. The
horizontal shaded region corresponds to the observed neutrino flux
at 25 TeV in the combined analysis
\citep{Aartsen2015ApJ}.}%
\end{figure}

\begin{figure}[ptb]
\includegraphics[width=0.9\textwidth]{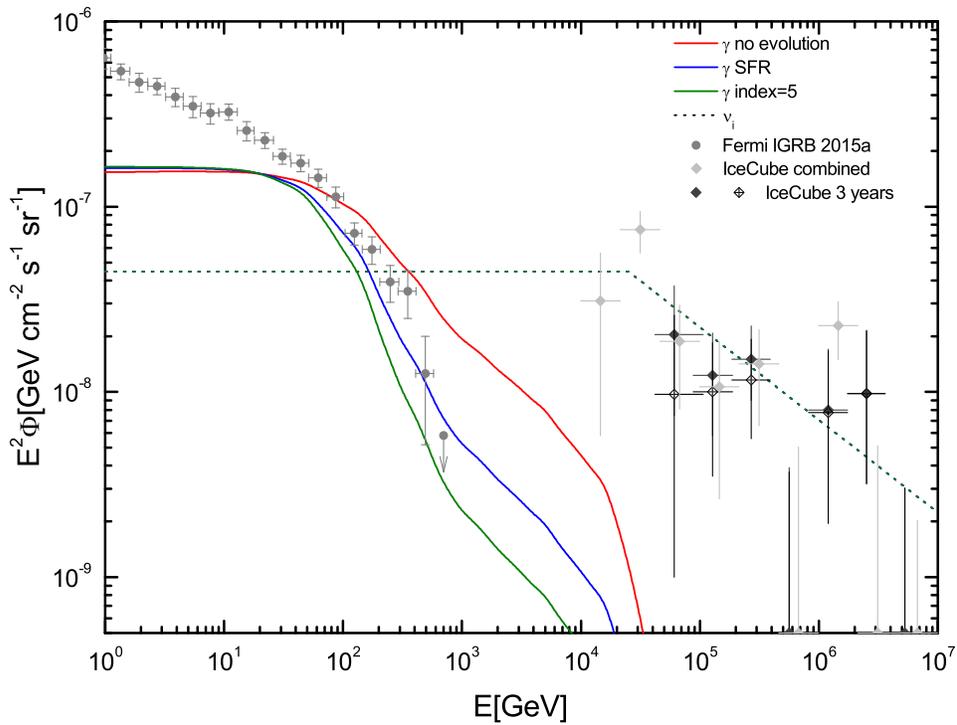}\caption{Comparison between the
cumulative gamma-ray emission of unresolved sources and the IGRB data for
different density evolution models. In the calculation, we use the case of
source density of $4\times10^{-4}\mathrm{Mpc^{-3}}$. The neutrino flux data
are denoted by black and grey dots. }%
\end{figure}

\begin{figure}[ptb]
\includegraphics[width=0.9\textwidth]{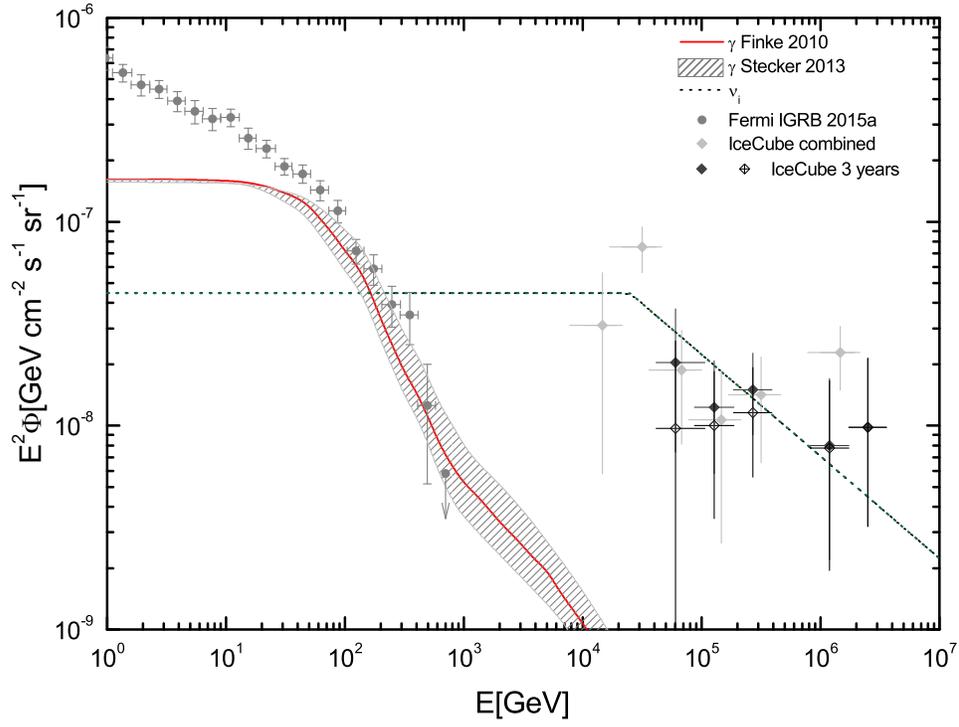}\caption{Same as figure 4, but
considering the uncertainty in EBL models. We use the upper and lower bounds
on the opacity given by \citet{Stecker2013arXiv} as the boundary of EBL
uncertainties. The red line shows the gamma-ray emission corresponding to the
EBL model provided by \citet{Finke2010ApJ}. In the calculation, we use a
source density of $4\times10^{-4}$Mpc$^{-3}$ and assume SFR redshift
evolution. }%
\end{figure}
\end{center}

\end{document}